\documentclass[prl,twocolumn,showpacs,preprintnumbers,amsmath,amssymb,nofootinbib,superscriptaddress,floatfix]{revtex4}
\usepackage{dcolumn}
\usepackage{bm}
\usepackage{graphicx}
\usepackage{color}
\usepackage{ifpdf}

\ifpdf
\else%
\fi

\def\epspdffile#1{
\ifpdf
  \scalebox{0.25}{\includegraphics[width=30cm]{#1.pdf}}
\else
  \scalebox{0.25}{\includegraphics[width=30cm]{#1.eps}}
\fi
}

\font\twelvebb=msbm12
\font\tenbb=msbm10
\font\sevenbb=msbm7
\newfam\bbfam

\textfont\bbfam=\twelvebb
\scriptfont\bbfam=\tenbb
\scriptscriptfont\bbfam=\sevenbb

\def\sqr#1#2{{\vcenter{\hrule height.#2pt
   \hbox{\vrule width.#2pt height#1pt \kern#1pt
      \vrule width.#2pt}
   \hrule height.#2pt}}}

\def\bsqr#1#2{{\vrule width #1pt height#2pt}}
\def\bsquare{{\mathchoice\bsqr66\bsqr66\bsqr33\bsqr33}}
\def\twovector[#1,#2]{\left(\begin{array}{c} #1 \\ #2 \end{array}\right)}
\def\badbreak{\penalty1000}

\def\rmsub#1#2{#1_{\mbox{\tiny #2}}} 
\def\mg{MG}

\begin{document}

\preprint{BUHEP-07-05}

\title{Adaptive Multigrid Algorithm for Lattice QCD}

\author{J. Brannick}
\affiliation{Department of Mathematics, The Pennsylvania State University,\\
  230 McAllister Building, University Park, PA 16802,  
  United States of America}
\author{R. C. Brower}
\affiliation{Center for Computational Sciences,  Boston University,\\
  3 Cummington Street, Boston,  MA 02215,  United States of America}
\affiliation{Department of Physics,  Boston University,\\
  590 Commonwealth Avenue, Boston,  MA 02215,  United States of America}
\author{M. A. Clark}
\affiliation{Center for Computational Sciences,  Boston University,\\
  3 Cummington Street, Boston,  MA 02215,  United States of America}
\author{J. C. Osborn}
\affiliation{Center for Computational Sciences,  Boston University,\\
  3 Cummington Street, Boston,  MA 02215,  United States of America}
\author{C. Rebbi}
\affiliation{Center for Computational Sciences,  Boston University,\\
  3 Cummington Street, Boston,  MA 02215,  United States of America}
\affiliation{Department of Physics,  Boston University,\\
  590 Commonwealth Avenue, Boston,  MA 02215,  United States of America}

\date{\today}

\begin{abstract}
  We present a new multigrid solver that is suitable for the Dirac operator
  in the presence of disordered gauge fields. The key behind the
  success of the algorithm is an adaptive projection onto the coarse
  grids that preserves the near null space. The resulting algorithm
  has weak dependence on the gauge coupling and exhibits very little
  critical slowing down in the chiral limit. Results are presented for
  the Wilson Dirac operator of the 2d U(1) Schwinger model.
\end{abstract}

\maketitle

The most demanding computational task in lattice QCD simulations
consists of the calculation of quark propagators, which are needed
both for generating gauge field configurations with the appropriate
measure and for the evaluation of most observables.  The calculation
of a quark propagator, which in the course of a simulation must be 
carried out innumerous times with varying sources and gauge backgrounds,
consists in turn of solving a very large system of linear equations,
\begin{equation}
D(U) \psi = \chi,
\label{eq:Dpsichi}
\end{equation}
where $\psi$ is the quark propagator, $\chi$ is the source term and
$D(U)$ is the discretized the Dirac operator matrix, with elements
dependent on the gauge field background $U$.

In the language of applied mathematics, Eq.~\ref{eq:Dpsichi} is a
discretized elliptic partial differential equation (PDE).  For
definiteness,
\begin{eqnarray*}
D_{x,y} = & - \frac{1}{2} \displaystyle \sum_{\mu=1}^{d} \bigl( &(1-\gamma_\mu)U_x^\mu\, 
\delta_{x+\hat\mu,y}\, +  \\
& & (1+\gamma_\mu)U_{x-\hat\mu}^{\mu \dagger}\, \delta_{x-\hat\mu,y}\bigr) + 
(2d + m)\delta_{x,y}
\end{eqnarray*}
is the discretized Dirac operator describing a fermion in \(d\)
dimensions with mass \(m\) in the Wilson discretization of the
Dirac equation.  In the full 4 dimensional QCD problem 
the matrices \(\gamma_\mu\) 
are the \(4\times4\) Dirac spin matrices and \(U\) is the \(SU(3)\) 
gauge field.  The Wilson discretization is not the only
one available, but it enters as a crucial
ingredient in the chirality preserving ``overlap'' and ``domain wall''  
discretizations \cite{Kaplan:1992bt,Shamir:1993zy,Neuberger:1998fp}.  
Moreover, many of the problems encountered in 
solving the Wilson-Dirac equation extend to other formulations,
such as the ``staggered'' fermion discretization. For these
reasons, in this paper we will concentrate on the Wilson
discretization.

For any realistic QCD simulation the size of the matrix in
Eq.~\ref{eq:Dpsichi} is too large for using a direct solver.
Iterative Krylov-space methods, made possible by the sparsity of the
matrix, must be used for calculating the propagators and very
efficient solvers have been developed.  Yet, as the system being
considered grows in size (for forefront simulation on a $64^4$
lattice, \(D(U)\) is a \(200M\times200M\) complex matrix) and the
quark mass in lattice units is brought toward zero, the condition
number of the matrix increases rapidly and so does the computational
cost of the solution.

In the field of applied mathematics it has been known for some time
that in such circumstances the separation of physical length scales
can be a very effective paradigm for improving the effectiveness of
numerical algorithms.  This paradigm has proven correct whether for
evolving Monte Carlo processes, modeling chemical reactions, or
molecular dynamics.  This is especially true when it comes to solving
systems of the form \(Ax=b\), where \(A\) is the sparse matrix that
arises from the discretization of continuum differential equations,
\(b\) is a source vector and \(x\) is the desired solution vector.
For such systems the multigrid (\mg) approach, where discretizations on
successively coarser grids are used to accelerate the solution finding
process, has proven to be the method to beat.

One exception to the above statement is in solving the Dirac operator
in lattice QCD: here the nature of the underlying gauge field in the
Dirac operator has proven to be especially resistant to
various \mg\ approaches.  Previous attempts at \mg\ solvers have
relied on renormalization group arguments to define the coarse grids
without realizing why the \mg\ approach succeeds, and this has
invariably led to failure as the physically interesting regime is
approached ~\cite{Brower:1991xv,Lauwers:1992cp}.  In this letter we
demonstrate a \mg\ algorithm for the Dirac operator normal equations,
i.e., the positive definite operator given by
\begin{displaymath}
A= D^\dagger D,
\end{displaymath}
that is shown to work in all regimes and vastly reduces the notorious
critical slowing of the solver as the renormalized fermion mass is
brought to zero.  We do so in the context of a 2-dimensional system
with $U(1)$ gauge field (Schwinger model).  This system captures many
of the physical properties (confinement, chiral symmetry breaking,
existence of non-trivial topological sectors) of the more complex
4-dimensional QCD.

The original formulation of \mg\ is best viewed with the example of
the free Dirac operator.  Multigrid solvers are based on the
observation that stationary iterative solvers (e.g., Jacobi,
Gauss-Seidel) are only effective at reducing local error components
leaving slow to converge, low wave-number components in the error.
For the free Dirac operator these slow modes will be smooth and can be
accurately represented on a coarser grid using simple linear
averaging.  However, on the coarse grid these low wave-number error
components become modes of shorter range and so relaxation should be
effective at removing them.  This process can continue, moving to
coarser and coarser grids until we have thinned the degrees of freedom
enough to solve the system exactly.  We then promote our solution back
to the finest grid, where at each level we relax on our correction
vector to remove any high wave-number error components that are
introduced.  This process is known as a \mg\
V-cycle~\cite{Brandt:1977} and can be used as a solver in its own
right, or more effectively as a preconditioner for a Krylov method
(e.g., conjugate gradient).

To help facilitate our discussion we introduce the notation where the
degree of coarseness is represented by the integer \(L\), where
\(L=1\) represents the finest grid (i.e., where our actual problem is
defined) and \(L=N\) is the coarsest grid in an \(N\)-level \mg\
algorithm.  The operator used to promote a coarse grid vector on grid
\(L=l+1\) to the adjacent fine grid \(L=l\) is known as the
prolongator \(P^{(l,l+1)}\), and it is convenient to take
\(P=P^\dagger\) as the restriction operator used for moving from the
fine grid to the coarse grid. (This guarantees Hermiticity of the
coarse grid operator in Eq.~\ref{eq:Galerkin}).  Typically the Galerkin
definition is used to define the coarse grid
operator~\cite{Brandt:1977},
\begin{equation}
  A^{(l+1)} = P^{(l,l+1)\dagger} A^{(l)} P^{(l,l+1)}.
\label{eq:Galerkin}
\end{equation}
That this is the best definition for Hermitian positive definite \(A\)
can easily be found by minimizing the error of the coarse grid
corrected solution vector in the \(A\)-norm.

Apart from the coarsest level which is just an exact solve, each level
of the \mg\ V-cycle can be succinctly described as
\begin{enumerate}
\item Relax on the input vector, \(x^{(l)} = R^{(l)\dagger} b^{(l)}\),
  where \(R^{(l)\dagger}\) is a suitable relaxation
  operator.~\footnote{The relaxation operator need not be Hermitian
    for the entire V-cycle to be Hermitian: the post-relaxation
    operator need only be the Hermitian cojugate to
    pre-relaxation.}
\item Restrict the resultant residual to the next coarsest grid, 
  \(r^{(l+1)} = P^{(l,l+1)\dagger} (b^{(l)}-A^{(l)}x^{(l)})\).
\item Apply the \(L=l+1\) V-cycle on the coarse residual, 
  \(e^{(l+1)} = V^{(l+1)} r^{(l+1)}\).
\item Correct the current solution with coarse grid correction,
  \(x^{(l)} = x^{(l)} + P^{(l,l+1)} e^{(l+1)}\).
\item Relax on the final residual, \( x^{(l)} = R^{(l)} (b^{(l)} - Ax^{(l)})\).
\end{enumerate}
Written explicitly in terms of operators the \(l^{th}\) level of the
V-cycle thus takes the following form
\begin{eqnarray}
V^{(l)} & = & R^{(l)} + R^{(l)\dagger} + R^{(l)}A^{(l)}R^{(l)\dagger} + \\\nonumber
& & \Big[ (1 - R^{(l)}A^{(l)}) P^{(l,l+1)} V^{(l+1)} \\\nonumber
& & P^{(l,l+1)\dagger} (1 - A^{(l)} R^{(l)\dagger}) \Big].
\label{eq:vcycle}
\end{eqnarray}
In this form the Hermiticity of the V-cycle is obvious.  The cost of
applying the \mg\ V-cycle becomes apparent from this explicit form: on
each level we must apply the operator \(A^{(l)}\) a total of \(2\nu +
2\) times for each \(l\), where \(\nu\) is the number of steps within
the relaxation operator.

The problem in the early application of the above procedure to the
interacting theory is that, in the presence of a non-trivial gauge
field, the eigenvectors responsible for slow convergence are no longer
low wave-number modes with smooth variation over the lattice.  They
are instead modes that exhibit localized lumps, typically extending
over several lattice spacings.  In such circumstances, trying to use
smooth components of the fermion field, defined through a suitable
gauge fixing or by some gauge covariant procedure, for the definition
of the prolongator is bound to produce only a limited advantage.  This
is the method that was followed, e.g., in
Ref.~\cite{Brower:1991xv,Lauwers:1992cp}, where some acceleration was
obtained but critical slowing down was not fully removed.

A breakthrough in the application of multiscale methods to more
complex problems, such as the one at hand, has occurred with the
discovery of adaptive \mg\ techniques~\cite{Brezina:2004,
Brannick:2006}.  In the adaptive algorithm one lets the \mg\ process
itself define the appropriate prolongator by an iterative procedure
which we now concisely describe.

In the first pass, one uses relaxation alone to solve the homogenous
problem \(Ae=0\) with a randomly chosen initial error vector.  After a
certain number, \(\nu\), of relaxation steps, the relaxation procedure, which we
symbolically represent by
\begin{equation}
e \to e^\prime=(I-\omega A)^\nu e \equiv (I- \omega D^\dagger D)^\nu e,
\label{eq:relax}
\end{equation}
produces an \(e^\prime\) that essentially belongs to the space spanned
by the slow modes, so $e^\prime$ is now used to define a first
approximation to the prolongator $P$.  One blocks the variables of the
original lattice into subsets, which we denote by $S_j$.  From
$e^\prime$ we construct the vectors $e^\prime_j$, which are identical
to $e^\prime$ within $S_j$ and $0$ outside $S_j$, and the vectors of
unit norm $v_{1j}=e'_j /\vert e'_j \vert$.  The extra ``1'' index in
$v_{1j}$ has been introduced for a discussion that follows.  The
prolongator $P^{1,2} \equiv P^{1,2}_{ij}$ which maps a vector
$\psi_j^{(2)}$ in the coarse lattice, indexed by $j$, to the original
lattice, where $i$ denotes collectively the site, spin and possible
internal symmetry indices, is then defined by
\begin{equation}
P^{1,2}_{i,j}=v_{1j,i},
\label{eq:P1}
\end{equation}
where we have made explicit the fine lattice indices of $v_{1j}$.

There are variations on how to block the fine lattice, i.e.,~how to
define the sets $S_j$.  In the so called ``algebraic adaptive MG'' one
partitions the fine lattice into subsets on the basis of the magnitude
of the matrix elements of $A$.  Since such matrix elements in lattice
gauge theories are typically of uniform magnitude, differing rather in
phase or, in a broader sense, in orientation within the space of gauge
transformations, we chose instead to partition the lattice
geometrically into fixed blocks of neighboring lattice sites,
specifically $4\times 4$ squares in our study of the Schwinger model.
Maintaining a regular lattice on coarse levels will allow more
efficient parallel code with exact load balancing.

Another refinement of the technique consists of applying a simple
Richardson iteration to the vectors $v_{1j}$ before defining the
prolongator.  The choice of damping parameter in this smoothing
procedure is chosen to minimize the condition number of the resulting
coarse grid operator.  The term ``smoothed aggregation'' is used for
this.  Thus our overall technique can be referred to as ``geometric
adaptive smoothly aggregated MG''.

We come now to the crux of the adaptive \mg\ method.  We use the
prolongator defined above (Eq.~\ref{eq:P1}) to implement a standard
\mg\ V-cycle and apply it, like relaxation before, to a randomly
chosen error vector.  There are two possibilities.  Either the V-cycle
reduces the error with no sign of critical slowing down or some large
error, $e^{\prime \prime}$, survives the cycle.  In the first case, of
course, one need not proceed: the \mg\ procedure works as is.  In the
second case, we define another set of vectors $v_{2j}$ over the coarse
lattice by restricting $e^{\prime \prime}$ to the subsets $S_j$,
making the new vectors orthogonal to the vectors $v_{1j}$ and
normalizing them to 1.  The smoothed aggregation procedure is now
applied to the set $v_{sj}\equiv (v_{1j}, v_{2j})$.  A new prolongator
is defined by projecting over these vectors
\begin{displaymath}
P_{i,sj}^{1,2}=v_{sj,i},
\end{displaymath}
where the index $s$, now taking values $1,2$, can be considered
as an intrinsic index over the coarse lattice.

The procedure described in the above paragraph is repeated as
necessary, until the application of a V-cycle reduces a random 
initial error to 0 without critical slowing down.  The method
works if critical slowing down is eliminated with a few iterations
of the adaptive procedure.   If this occurs with $M$ vector sets,
then the coarse lattice will carry $M$ degrees of freedom per site.
As with all \mg\ methods, the procedure is recursive and it can
be used to define further coarsenings.

In testing this algorithm for lattice QCD we generated quenched $U(1)$
gauge field configurations on a $128 \times 128$ lattice with the
standard Wilson gauge field action
\begin{displaymath}
S=\sum_{x,\nu<\mu} \beta\, {\rm Re}\, U_x^{\mu \nu} \equiv
\sum_{x,\nu<\mu} \beta\, {\rm Re}\, U_x^{\mu} U_{x+\hat \mu}^{\nu}
U_{x+\hat \nu}^{\mu \dagger} U_x^{\nu \dagger}
\end{displaymath}
and periodic boundary conditions at $\beta=6$ and $\beta=10$ at a wide
range of mass parameters.  These two values of $\beta$ define
correlation lengths for the gauge field to be $l_\sigma = 3.30 $ and
$l_\sigma = 4.35$ respectively, via the area law for the Wilson loop: $
W \sim \exp[ -A/l^2_\sigma]$. For comparison on these lattices, a
fermion mass gap \(\hat{m} = m - \rmsub{m}{crit} = 0.01\) corresponds
to the pseudoscalar meson correlation lengths \(\mu^{-1} = 6.4 \) and
\(\mu^{-1} = 12.7 \) respectively.~\footnote{All quantities are
  expressed in lattice units.}  In the 2-dimensional \(U(1)\) gauge
theory, one can identify a gauge invariant topological charge \(Q\),
which in the continuum limit is proportional to the quantized magnetic
flux flowing through the system.  A gauge field with nonzero \(Q\)
corresponds to a Dirac operator with exactly real eigenvalues and,
hence, as the mass gap is brought towards zero the condition number
becomes infinite.  Thus, it is important to test both trivial
(\(Q=0\)) and non-trivial (\(Q\neq 0\)) gauge field topologies.

We blocked the lattice into $4 \times 4$ blocks and implemented the
adaptive smoothly aggregated \mg\ procedure described above.  We used
a degree 2 polynomial smoother for our relaxation procedure, where the
coefficients were chosen by running two iterations of an underrelaxed
minimum residual solver (\(\omega=0.8\)) and subsequently held fixed
(hence, for our choice of smoother \(R=R^\dagger\)).  The coarsening
procedure was repeated twice maintaining $M=8$ vectors in all
coarsenings, down to an $8 \times 8$ lattice, over which the equations
were solved exactly.  For each gauge field we performed the set up
procedure for the \mg\ preconditioner for the lightest mass parameter
only, and reused these null space vectors for the heavier masses. We
used this constructed V-cycle as a preconditioner for the conjugate
gradient (CG) technique where the operator defined in
Eq.~\ref{eq:vcycle} is applied at each iteration to the CG direction
vector.

\begin{figure}[htb]
\epspdffile{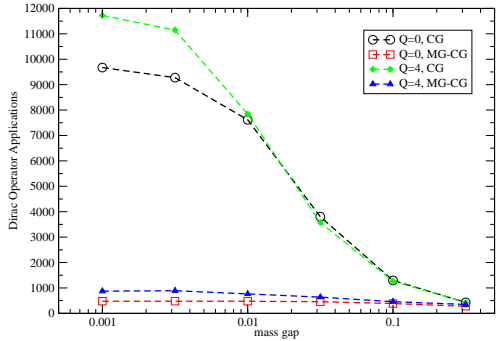}
\vspace{-7mm}
\caption{\label{fig:beta6}Number of Dirac operator applications of
  standard CG vs.~MG-preconditioned CG solver as function of the
  fermion mass gap at $\beta =6$ with topological number $Q=0$ and
  $Q=4$ (point source, relative solver residual \(|r|=10^{-14}\)).}
\end{figure}
\begin{figure}[htb]
\epspdffile{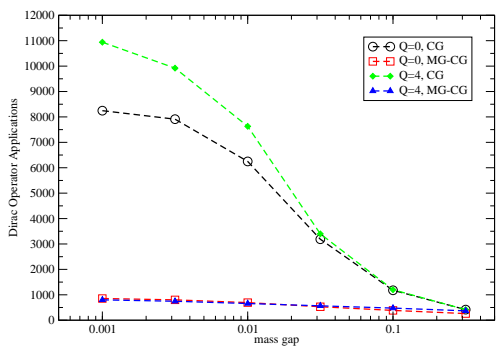}
\vspace{-7mm}
\caption{\label{fig:beta10}Number of Dirac operator applications of
  standard CG vs.~MG-preconditioned CG solver as function of the
  fermion mass gap at $\beta =10$ with topological number $Q=0$ and
  $Q=4$ (point source, relative solver residual \(|r|=10^{-14}\)).}
\end{figure}

If one compares the number of CG iterations needed to achieve
convergence with or without \mg\ preconditioning, the gain obtained
with the \mg\ method is dramatic: for example, with \(\beta =6\),
\(\hat{m}=0.01\) and \(Q=0\), it takes 3808 iterations to achieve
convergence, in the sense above, with a straightforward application of
the CG technique, whereas it takes only 26 iterations using the \mg\
preconditioner.  However this comparison does not take into account
the fact that many more operations per iterations must be performed
when applying the \mg\ preconditioner.  To achieve a more balanced
comparison, in Figs.~\ref{fig:beta6},~\ref{fig:beta10} we plot the
total number of applications of \(D\) and \(D^\dagger\) done on the
fine lattice.  This reflects better the actual cost of the
calculations (at each iteration of MG-CG there are 6 applications of
\(D^\dagger D\): 1 application in the outer CG, and 2 pre- and 2 post-
coarsening smoothing applications and 1 further application required
to form the residual).  We do not include the additional cost arising
from the coarse lattices since this is expected to be a small
overhead, and has not been optimized for our model calculation.  The
advantage coming from the use of the adaptive \mg\ technique is still
very dramatic.  In particular, we see that critical slowing down, if
not totally eliminated, is very substantially reduced.  These results
are for point sources, however, we tried a variety of different source
vectors for this analysis (e.g., Gaussian noise, \(Z_4\) noise) and
found very little dependence of MG-CG performance on the source
vector.

From the point of view of computational complexity, one should also
take into account the cost of setting up the \mg\ preconditioner,
i.e.,~of constructing the prolongator $P$.  This cost is however
heavily amortized, to the point of being negligible, if, as is often
the case, one must apply the solver to systems with multiple given
vectors (for example, solving for all color and spin components of a
quark propagator or, in the calculation of disconnected diagrams where,
\(O(1000)\) inverses are required to estimate the trace of the inverse
Dirac operator).

Our results, albeit for now limited to a 2-dimensional example,
provide a clear indication that adaptive \mg\ can be made to work with
the lattice Dirac operator.  What appears to be at the root of its
success is that, although the modes responsible for slow convergence
of the Dirac solver on a fine lattice are not low wavenumber
excitations, like in the free case, their span can be well
approximated by a set of vectors of limited dimensionality on the
blocks that define the coarse lattice.  Earlier
attempts~\cite{Brower:1991xv,Lauwers:1992cp} tried to find the
approximating subspaces on the basis of smoothness, failing to
eliminate critical slowing down when the pseudoscalar length exceeded
the disorder length of the gauge field: $\mu^{-1} > l_\sigma$.
Adaptive \mg\ finds the coarse subspaces through the iterative
application of the method itself.  It is of course crucial that the
approximation to the space of slow modes can be achieved with a small
number of vectors on the individual blocks, otherwise the application
of the method would not be cost effective.  But this appears to be the
case in the examples we studied and, if the results hold true in
general, adaptive \mg\ has the potential of substantially speeding up
lattice QCD simulations as the increase of available computational
power leads one to consider ever larger lattices.  The observation
that the space of slow modes may be of limited span is also at the
root of a method recently proposed by L\"uscher in
Ref.~\cite{Luscher:2007se}, although the technique there is quite
different from the one we follow.  The application of the method to
4-dimensional systems is in progress.

Acknowledgments.  This research was supported in part under
DOE grants DE-FG02-91ER40676 and DE-FC02-06ER41440 and NSF
grant PHY-0427646.


\end{document}